\documentclass[apj]{emulateapj}
\usepackage{epsf,graphicx}
\usepackage{hyperref}
\def\la{\mathrel{\hbox{\rlap{\hbox{\lower4pt\hbox{$\sim$}}}\hbox{$<$}}}}
\def\ga{\mathrel{\hbox{\rlap{\hbox{\lower4pt\hbox{$\sim$}}}\hbox{$>$}}}}

\def\ltsima{$\; \buildrel < \over \sim \;$}
\def\lsim{\lower.5ex\hbox{\ltsima}}
\def\gtsima{$\; \buildrel > \over \sim \;$}
\def\gsim{\lower.5ex\hbox{\gtsima}}

\def\eg{{e.g.}}
\def\ie{{i.e.}}

\def\phix{\ifmmode \Phi_{\rm max} \else $\Phi_{\rm max}$\fi}
\def\phin{\ifmmode \Phi_{\rm min} \else $\Phi_{\rm min}$\fi}
\def\nx{\ifmmode n_{\rm max} \else $n_{\rm max}$\fi}
\def\nn{\ifmmode n_{\rm min} \else $n_{\rm min}$\fi}

\def\ll{\ifmmode \lambda \else $\lambda$\fi} 

\def\un{\ifmmode U_{\rm min} \else $U_{\rm min}$\fi}
\def\Un{\ifmmode U_{\rm min} \else $U_{\rm min}$\fi}

\def\phih{\ifmmode \Phi_{\textrm{\tiny H}} \else $\Phi_{\textrm{\tiny H}}$\fi}
\def\nh{\ifmmode n_{\textrm{\tiny H}} \else $n_{\textrm{\tiny H}}$\fi}
\def\Nh{\ifmmode N_{\textrm{\tiny H}} \else $N_{\textrm{\tiny H}}$\fi }

\def\lya{\ifmmode {\rm Ly}\alpha \else Ly$\alpha$\fi}
\def\lyb{\ifmmode {\rm Ly}\beta \else Ly$\beta$\fi}
\def\pa{\ifmmode {\rm Pa}\,\alpha \else Pa$\,\alpha$\fi}
\def\pb{\ifmmode {\rm Pa}\,\beta \else Pa$\,\beta$\fi}   
\def\pc{Pa$\,\gamma$}

\def\Ha{\ifmmode {\rm H}\alpha \else H$\alpha$\fi}
\def\Hb{\ifmmode {\rm H}\beta \else H$\beta$\fi}
\def\Hc{\ifmmode {\rm H}\gamma \else H$\gamma$\fi}

\def\HI{H$\,${\sc i}}

\def\HeI{He$\,${\sc i}}

\def\CaII{Ca\,{\sc ii}}      		
\def\FeII{Fe\,{\sc ii}}      		
\def\MgII{Mg\,{\sc ii}}      		
\def\NaI{Na\,{\sc i}}        		
\def\CIV{C\,{\sc iv}}     		
\def\OVI{O\,{\sc vi}}     		
     		
\def\CIII{C\,{\sc iii}]}     		
\def\Caxyz{Ca\,{\sc ii\,xyz}}     		
\def\cloudy{{Cloudy}}
\def\ew{\ifmmode W_{1216} \else $W_{1216}$\fi}
\def\ewl{\ifmmode W_{\lambda} \else $W_{\lambda}$\fi}
\def\ewopt{\ifmmode W_{5100} \else $W_{5100}$\fi}
\def\fc{\ifmmode f_{c} \else $f_{c}$\fi}
\def\kms{\ifmmode \,{\rm km}\,{\rm s}^{-1} \else $\,$km$\,$s$^{-1}$\fi}
\def\cmsq{\ifmmode \,{\rm cm}^{-2} \else $\,$cm$^{-2}$\fi}
\def\cmsqs{\ifmmode \,{\rm cm}^{-2}\,{\rm s}^{-1} \else $\,$cm$^{-2}\,$s$^{-1}$\fi}
\def\cmcu{\ifmmode \,{\rm cm}^{-3} \else $\,$cm$^{-3}$\fi}
\def\micron{\ifmmode \mu \else $\mu$\fi}

\def\objone{J005812$+$160201}
\def\objtwo{J010226$-$003904}

\def\objfive{J015950$+$002340}
 
\def\objseven{J032213$+$005513}

\def\galex{GALEX}
\def\sdss{SDSS}

\newdimen\digitwidth
\setbox0=\hbox{-.}
\digitwidth=\wd0
\catcode `@=\active
\def@{\kern\digitwidth}

\shorttitle{New Constraints on the Quasar Broad Emission Line Region}
\shortauthors{Ruff et~al.}

\begin{document}

\title{New Constraints on the Quasar Broad Emission Line Region}

\author{Andrea~J.~Ruff\altaffilmark{1}$^{*}$}
\author{David~J.~E.~Floyd\altaffilmark{1,2}}
\author{Rachel~L.~Webster\altaffilmark{1}}
\author{Kirk~T.~Korista\altaffilmark{3}}
\author{Hermine~Landt\altaffilmark{1}}

\altaffiltext{1}{School of Physics, University of Melbourne, Parkville, 
VIC 3010, Australia.}
\altaffiltext{2}{School of Physics, Monash University, VIC 3800, Australia}
\altaffiltext{3}{Department of Physics, Western Michigan University, 
Kalamazoo, MI 49008.}

\altaffiltext{*}{{\tt aruff{@}unimelb.edu.au}} 


\begin{abstract}

We demonstrate a new technique for determining the physical conditions of the
broad line emitting gas in quasars, using near-infrared hydrogen emission
lines. Unlike higher ionisation species, hydrogen is an efficient line emitter
for a very wide range of photoionisation conditions, and the observed line
ratios depend strongly on the density and photoionisation state of the gas
present.  A locally optimally emitting cloud model of the broad emission line 
region was compared to measured emission lines of four nearby
($z\approx0.2$) quasars that have optical and NIR spectra of sufficient
signal-to-noise to measure their Paschen lines. 
The model provides a good fit to three of the objects, and a fair fit to the
fourth object, an ultraluminous infrared galaxy.  We find that low incident ionising fluxes
($\phih<10^{18}$\cmsqs), and high gas densities ($\nh>10^{12}$\cmcu) are
required to reproduce the observed hydrogen emission line ratios.  This analysis demonstrates
that the use of composite spectra in photoionisation modelling is
inappropriate; models must be fitted to the individual spectra of quasars.

\end{abstract}



\section{Introduction}\label{section:intro}

Physical models of the broad emission line region (BELR) of active galactic
nuclei (AGNs) have been frustrated by the complexity of the expected emission
mechanisms, and a poor understanding of the physical state, dynamics and
spatial distribution of the line emitting gas.  Innovative new techniques,
including reverberation mapping \citep{peterson1994_review} and microlensing
\citep[\eg][]{odowd2011,bate2008}, are providing indirect methods to measure
the physical sizes of the emission regions. To date, each of these techniques
has only been applied to a select sample of sources.  However, useful models of
the BELR will require constraints on all physical parameters.  Broad
emission line fluxes provide a powerful diagnostic of the physical conditions
in the BELR, particularly when coupled with photoionisation simulations from
codes, such as \cloudy.

Broad emission lines are a prominent feature in luminous AGNs, and a
defining characteristic of quasars.  These emission lines are typically
broadened by gas velocities of $\sim$10$^4$\kms, and are understood to arise
from gas close to the central black hole and its accretion disk. The broad
emission line gas reprocesses a substantial amount of the light emitted from
the central region. The emitted line fluxes, and in particular their ratios,
are strongly dependent on the local physical conditions.  
The broad emission line region is the closest material that interacts strongly
with the accretion disk that can be probed observationally, and a detailed
physical understanding is essential in understanding the role of AGN
energy injection at larger scales.

A compact broad emission line region has been inferred from
reverberation mapping \citep[\eg,][]{blandford1982_reverb,peterson1994_review}.
As a result of the small spatial scale of the BELR and large distances to the
nearest quasars, direct spatial imaging of the BELR is not possible. 
In the absence of detailed structural
information, we can develop photoionization models to predict spectral line
ratios and infer the properties of the emitting gas. Such calculations are
particularly sensitive to the density and ionisation state of the
gas~\citep[\eg][]{davidson1972,osterbrock1986}.  This provides an alternative
estimate of the physical scale of the BELR.

Early BELR photoionisation simulations were limited to single cloud models with
a single number density, column density and ionisation parameter. Simulating the entire range of possible BELR
parameter space was not possible, due to
computational limitations. These simulations were reasonably successful
in reproducing a number of observed emission line flux ratios
\citep[\eg][]{davidson1977,kwan1981}. However, as more spectroscopic data
became available, there was conflicting evidence: on one hand, the presence of
semi-forbidden transitions, such as \CIII\ \ll1909 suggested reasonably low
number density gas \citep{osterbrock1970}, while the hydrogen emission line
ratios and strong \FeII\ emission suggested the presence of high number density
gas \citep[\eg][]{rees1989,collinsouffrin1982_hydrogen}. 
%
%
Broad emission line-continuum reverberation measurements find a wide range
in emission line lags relative to the continuum variations and line widths,
arguing for a distributed geometry of physical conditions \citep[\eg, see][]{krolik1991,peterson2011}.
To reconcile these conflicting observations, \citet{baldwin1995_loc} proposed a
locally optimally emitting cloud (LOC) model, where gas `clouds' with a range
of physical properties exist at each radius in a spatially extended BELR. This
model takes advantage of large grids of photoionisation models that can now be
simulated over a huge range of parameter space in the density-flux plane
\citep{korista1997_atlas}. 

The LOC model has been successful in producing ultraviolet and optical emission
line flux ratios that match observations
\citep[][]{baldwin1995_loc,KG2000,bottorff2000}, while maintaining consistency
with reverberation mapping results \citep[\eg][]{KG2004}. The LOC also provides
an explanation for the remarkably consistent BELR flux ratios, over a very
large range in AGN luminosity.  The name, `LOC' originated because many of the
optical and ultraviolet (UV) broad emission lines emit `optimally' in one region
in parameter space. 
%
%
Specifically, the typical emission line is efficiently emitted (relative to the
incident continuum) over a relatively narrow region within the density-incident
continuum flux plane \citep[see Fig. 1 of][]{baldwin1995_loc}. This natural
selection effect is due to a combination of ionization potential, collisional
de-excitation of the upper level, and thermalization at large optical depths. 
Reverberation mapping measurements show that individual emission lines respond to variations in the
continuum with different time delays, and the characteristic radius where each
line is produced depends on the ionisation potential (\ie\ high ionisation
lines are emitted closer to the continuum source).
The term, `LOC,' is less appropriate for the \HI\ and \HeI\ lines, as the
production of these lines from the continuum remains efficient over a large
range of the density-flux plane \citep{korista1997_atlas,ruff2011b}. 

The hydrogen and helium lines are of particular interest because the physical
processes governing their production is relatively simple, and there is a
negligible dependence on several free parameters of the model: metallicity,
column density and slope of the ionising continuum.
Despite the relative simplicity of the \HI\ emission spectrum, some unexplained
observations persist, such as the \lya/\Hb\ problem
\citep{baldwin1977,netzer1995} and the small scatter in measured Balmer
decrements (\Ha/\Hb) of AGNs with blue continua \citep{dong2008}. 
%
%
The simple recombination spectrum predictions of Case~B are invalid for several
reasons under BELR conditions. Firstly, given the high gas number density,
high incident flux environment, the optical depths in the Balmer and Paschen
lines are finite and often large. 
Secondly, the effects of collisions  are expected to be very important,
especially in light of the enhanced lifetime of the $n=2$ state due to the
enormous \lya\ optical depths that are encountered.

%

Although the LOC model has been shown to produce emission line flux ratios that
are consistent with measured emission lines for optical and UV metal emission
lines \citep{baldwin1995_loc,KG2000} as well as helium emission lines
\citep{bottorff2002,KG2004}, no comprehensive study of the infrared hydrogen
lines has been undertaken.  
The near-infrared (NIR) lines, for example the hydrogen Paschen series, have
the advantage of being affected by dust equally, whereas there is differential
extinction across the optical and UV lines. 
With advances in infrared spectroscopy, larger samples of both low luminosity
AGN \citep{landt2008} and higher luminosity quasars \citep{glikman2006} are
being observed. 
The Paschen lines are becoming more widely used for AGN measurements, such as
black hole mass \citep[\eg][]{kim2010}. 


This paper investigates a new diagnostic that tests the LOC model using
near-infrared hydrogen emission lines in quasars. The narrow component has a
larger contribution to the total line flux in lower luminosity AGNs, and the
systematic uncertainties resulting from this subtraction must be quantified. We
have concentrated on quasars (as opposed to Seyferts and other lower-luminosity
AGN), so that deconvolution of the emission lines into broad and narrow
components is not necessary.  We use only spectra of individual objects, and
demonstrate that composite spectra are not suitable for determining the
physical conditions in quasars.
Section~\ref{section:models} describes the photoionisation simulations and the
simple LOC model used in this analysis.  
Section~\ref{sect:data} describes data used. 
Measurements are compared to the LOC model in
Section~\ref{section:results}. A discussion of the results is given in
Section~\ref{section:discussion}, before a brief summary in
Section~\ref{section:summary}.


\section{Photoionization models}\label{section:models}

Following the standard LOC approach
\citep[]{baldwin1995_loc,korista1997_atlas,KG2000}, a large grid of
photoionisation models was computed using the photoionisation code, \cloudy\
version 08.00, last described by \citet{ferland1998}.  Version 08.00 of
\cloudy\ implemented full resolution of quantum $l$-levels\footnote{The
azimuthal quantum number, $l$, is one of the numbers that describes the quantum
state of an electron.} for the H and He-like sequences for the first time
\citep[][]{ferguson1997,ferguson2001,porter2005,porter2007}, greatly improving
the accuracy of the predicted emission. 
For greater predictive accuracy, the number of resolved quantum levels in the
\HI\ and \HeI\ atoms was increased from the \cloudy\
defaults\footnote{Simulations used \HI\ and \HeI\ model atoms with 18 and 15
resolved levels, respectively.}. A brief outline of the grid simulations is
given below. A more detailed description of the photoionisation simulations is
given by \citet{ruff2011b}. 

In the context of an LOC model, the term, ``cloud'' refers to an individual
emitting body of gas within the BELR. 
Emission line profiles are observed to be very smooth, implying either a large
number of clouds \citep{arav1998_clouds}, or a smoothing mechanism. 
Micro-turbulence of $\sim$100\kms\ would broaden the local line profile
sufficiently to smooth the bulk line profile to match the observed (\eg, as
mentioned in \citealp{arav1998_clouds}; see also
\citealp{bottorff2000,kol2011}). Micro-turbulence also reproduces the observed
UV \FeII\ bump \citep{baldwin2004_fe2}.  The presence of a small amount of
electron scattering \citep[\eg][]{emmering1992,laor2006} or macro-turbulence in
the cloud velocity field could also smooth the profile. 
%
%
%
The precise nature of the line-emitting entities is not the focus of this
investigation. Instead we focus on constraining general physical conditions of
the emitting gas using hydrogen emission line flux ratios.


 

%

\subsection{Grid simulations in the density-flux plane}\label{section:grid}

Models that predict hydrogen and helium emission line strengths were generated
for a range of hydrogen number density ($\nh$) and hydrogen ionizing flux
($\phih$) values. The plane formed by these parameters represents the range of
cloud densities and distances from the ionizing continuum that are expected to
exist within the BELR.  
The output emission of the photoionisation simulations is computed as an
equivalent width in angstroms relative to the incident continuum at 1216\AA\
($W_{1216}$). The equivalent width is a useful description of the line
emission, as it measures how efficiently the line is produced from the incident
continuum. 

For each point in the grid, \cloudy's standard AGN ionising continuum (similar
to the \citet{mathews1987_cont} continuum) was used. Despite a range in
observed ionising continua, the H and He emission lines are relatively
insensitive to the shape of the ionising continuum, unlike the metal emission lines
\citep{korista1997_atlas}. 
A constant number density within each cloud was assumed and a constant column
density of $\Nh$$\,=\,$$10^{23}\,$cm$^{-2}$ was also assumed. 
Although a range of column densities is expected within the BELR, the hydrogen
and helium emission lines are not very sensitive to the cloud column density in
the range $10^{22}$$\,<\,$$\Nh$(cm$^{-2}$)$\,<\,$$10^{24}$
\citep{korista1997_atlas,ruff2011b}. 
Note that solar chemical abundances were used for all photoionisation simulations. 

\subsection{Limits on the number density, ionising flux and ionisation parameter}\label{sect:param}

The range of values for each of the physical parameters in the LOC model is
motivated by observations and physical considerations. 
Broad, semi-forbidden \CIII\ $\lambda$1909 is observed, implying the presence of
$\sim$$10^9$\cmcu\ gas \citep{davidson1979}, as it is collisionally suppressed
at higher densities. 
However, broad forbidden lines are not observed, indicating that the electron
density in the BELR is higher than the critical density of the forbidden lines.
This gives a lower limit on the number density within the BELR and 
 only clouds with \nh$\,>\,$10$^7$\cmcu\ were considered. While none of the
broad emission lines can be used to determine an upper limit on the number
density, analyses of the Balmer lines, whose ratios are sensitive to high
density gas, and strong \FeII\ emission indicate that high density
($\sim$10$^{12}$\cmcu) gas is also present within the BELR
\citep[\eg,][]{collinsouffrin1982_hydrogen,rees1989}.
At these high densities, collisional excitation becomes important. The
collisional excitation depends on density, temperature and ionisation state in
a fundamentally different manner to the recombination Case~B
\citep{ferland2009_collisional}. Therefore detailed photoionization simulations are required.  

%
Reverberation mapping measurements show that broad emission lines originate at
a range of radii from the central ionising source \citep{peterson1994_review}.
At large distances from the continuum source, the ionising flux is low and the
temperature decreases. Below temperatures of $\sim$1500$\,$K dust grains
condense, making the line emissivity low \citep[\eg][]{netzer1993}. 
%
%
This temperature corresponds to a hydrogen ionising flux of 
$\phih$\,$\sim$\,$10^{18}$\cmsqs\ \citep[][after substituting a fiducial bolometric
luminosity]{nenkova2008b}.  The \HI\ and \HeI\ emissivities are strongly
dependent on the incident ionising flux, unlike most of the prominent metal
lines \citep{KG2004}. This is because the optical depth of the \HI\ and \HeI\
excited states increases with ionising flux
\citep{ferland1979b,ferland1992}, until the cloud is fully ionised. Close to
the ionising source, the hydrogen and helium atoms will be fully ionised and
will not produce line emission.  As the \HI\ and \HeI\ line emissivity is low
at large \phih\ values, the upper limit on $\phih$ is considered to be
unimportant. 

Considering the limits discussed above, a large range in
BELR parameter space was simulated, spanning seven decades in both \nh\ and \phih:
$10^{7}$$\,\leq$$\,\nh\,$(cm$^{-3}$)$\,\leq\,$10$^{14}$ and
$10^{17}\,$$\leq$$\,\phih\,$(cm$^{-2}\,$s$^{-1}$)$\,\leq\,$$10^{24}$,
each stepped in 0.25 decade intervals. Each model was calculated
separately by \cloudy, giving a total of 841 photoionization
calculations. Contour plots of each line, showing the \ew\ as a
function of \nh\ and \phih\ will be presented in \citet{ruff2011b}.
%
%
%
A lower limit on the ionisation parameter, \begin{equation} U\equiv
\frac{\phih}{c\,\nh} \end{equation} 
was also considered. This is motivated by observations of the calcium 
infrared triplet, \Caxyz\ \ll\ll\ll 8498, 8542, 8662. If clouds with very low values of the ionisation 
parameter ($\log U\leq -6$) existed in the BELR, stronger
\Caxyz\ \citep{joly1989,ferlandpersson1989} and \NaI\ \ll5895 emission would
be observed \citep[see][for a range of \ew\ distributions in the
density-flux plane]{korista1997_atlas}. In addition, the \FeII\
emission would be totally dominated by resonance transitions
clustered near 2400\AA\ and 2600\AA\ \citep{verner1999_fe2}, which is
dissimilar to the pseudo-continuum observed.
A physical interpretation of the lower limit on the ionisation parameter, \Un\ is that very
dense clouds are not found far from the ionising continuum. 
Although this limit may not exactly represent the true
physical cloud distribution, it is a convenient representation in the
LOC model.

These considerations provide only weak constraints on the upper \nh\ limit, and
lower \phih\ and $\Un$ limits in the BELR. However, they are very important
because of the large emissivity of \HI\ and \HeI\ close to these limits.  The
\HI\ and \HeI\ line ratios are particularly sensitive to the lower \phih\ and
upper \nh\ values (\phin\ and \nx, respectively). Thus rather than using
fiducial values, a range of upper and lower limits on \phih\ and \nh\ was
considered. 


\subsection{Calculating the total equivalent width}\label{sec:integration}

Following the standard LOC prescription, the flux from each point in the
simulated parameter space was summed to calculate the emission from each species
of interest. 
As shown by \citet{baldwin1995_loc}, the line luminosity is given by:
\begin{equation} 
L_{\rm line} \propto \int^{r_{\rm max}}_{r_{\rm
min}}\!\!\int^{n_{\rm max}}_{n_{\rm min}} \ew(r,\nh)\,f(r)\,g(\nh)\,d\nh\,dr, 
\label{equation:loc1}
\end{equation}
where $\ew(r,\nh)$ is the equivalent width of a particular line (relative to
the continuum at 1216\AA) from a single cloud at radius $r$, with number
density $\nh$. The cloud covering fractions as functions of  radius and number
density are given by $f(r)$ and $g(\nh)$, respectively.  For a more complete
description of the LOC parameters, see \citet{bottorff2002}.  In the simplest
case of LOC integration, the covering
%
%
fractions are given by $f(r) \propto r^{-1}$ and $g(\nh) \propto \nh^{-1}$ 
(\citealp{baldwin1995_loc}; see also \citealp{matsuoka2007}).
The line luminosity can then be simplified to: 
\begin{equation} L_{\rm line} \propto \int^{\log \Phi_{\rm max}}_{\log
\Phi_{\rm min}}\!\!\int^{\log n_{\rm max}}_{\log n_{\rm min}} \ew(\phih,\nh)\,
d\log \nh\,\,d\log \phih, 
\label{equation:loc}
\end{equation} 
%
where the integration limits are free parameters of the model and $\phih \propto
r^{-2}$.  
This is simply a sum over each point in parameter space between the integration
limits, since the simulated photoionized clouds in the logarithmic grid are
weighted evenly per decade. We note that this weighting scheme maximizes the
distribution of cloud characteristics in gas density and incident ionizing flux
-- a key characteristic of the LOC model. Power law indices in the density and
distance weighting functions $(f(r), g(\nh))$ which deviate substantially from -1
will take on characteristics of a model with a single gas density and ionizing
continuum source distance\footnote{In studies of the UV emission lines in
NGC~5548 \citep{KG2000} and SDSS quasars \citep{nagao2006} there is
evidence for favoring adoption of a slightly steeper power law index in $f(r)$,
which tends to favor gas at smaller distances from the central
continuum source. Here, however, given the limited number of observational
constraints (discussed in Section~\ref{sect:data}), we choose to adopt the simplest weighting
scheme in the LOC integration.}.

This sum (Equation~\ref{equation:loc}) then allows us to compute the predicted
total equivalent width (\ew) for each emission line, assuming full geometric
coverage. The model integrated covering fraction is then determined from the
ratio between the observed \ew(\lya) and its predicted value, above. While in
the present work we are mainly interested in comparing hydrogen Balmer and
Paschen emission line flux ratios (which are independent of the integrated
covering fraction), Table~\ref{table:predictew} lists the value of the
integrated covering fraction that results in a predicted equivalent width of
\lya\ of 100\AA, a typical value (see also
Section~\ref{section:ewsanitycheck}).

%
%


\subsection{Ratios calculated from varying the integration limits}\label{sect:limits}

\begin{figure}
\centering
\includegraphics[width=0.97\linewidth]{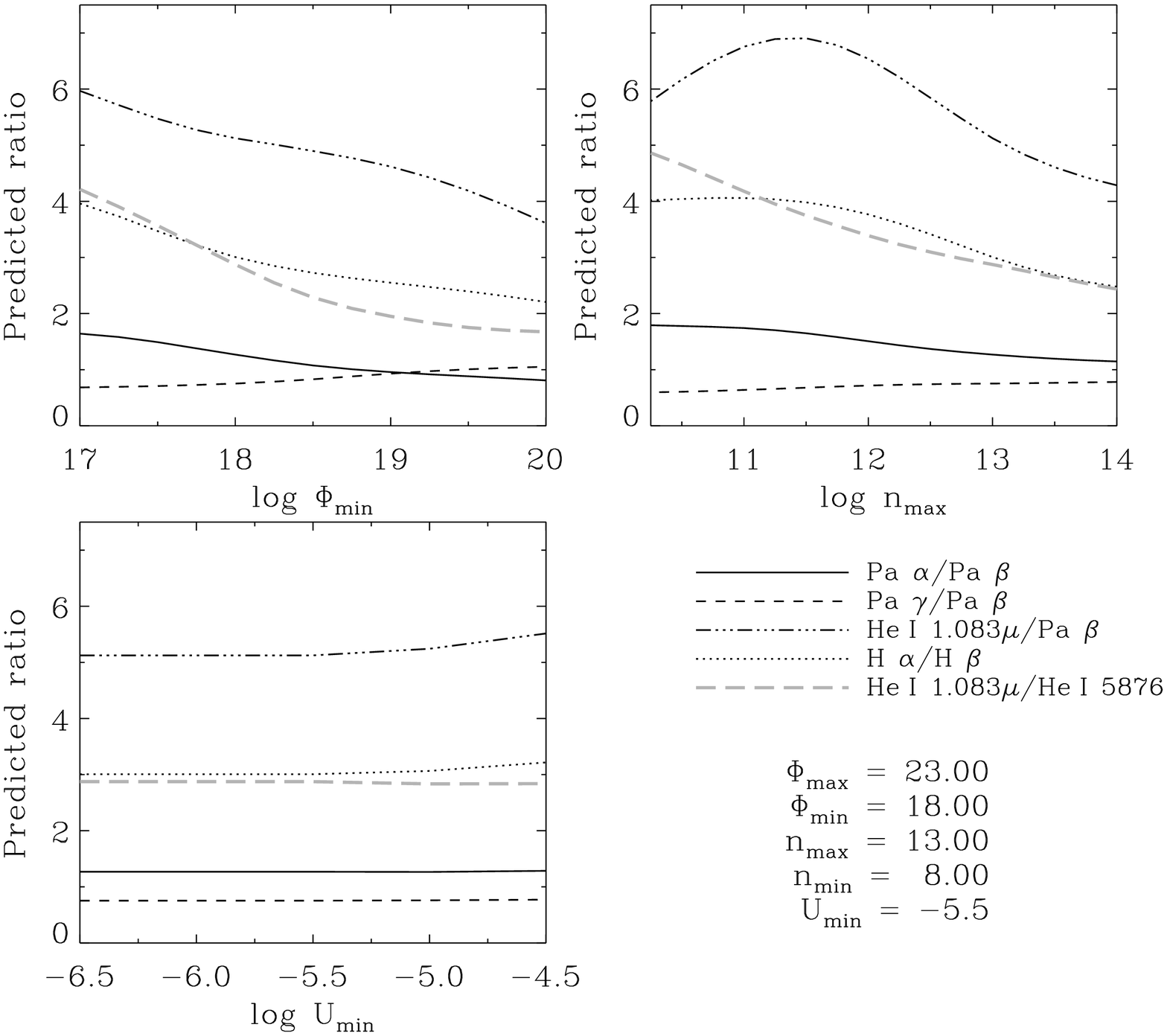}\\
\caption{\label{figure:varylims} 
Flux ratios are shown as a function of varying three integration limits: 
\phin, \nx\ and $\Un$. 
Each panel shows the variation of a different limit, while 
the other limits are held at constant at fiducial values, given in the legend. 
}
\end{figure}

Broad emission line flux ratios were predicted for several emission line pairs using Equation~\ref{equation:loc}. 
Rather than using fiducial values of the integration limits to predict ratios, 
a range of
upper and lower bounds on each of the free parameters: \phin, \nx\ and $\Un$
was considered.  
From the discussion in Section~\ref{sect:param},  the $\phin$, $\nx$ and $U_{\rm min}$ integration limits 
were varied over the ranges: 
\begin{equation} 
\label{eqn:intlimits}
\begin{array}{llc}
\log \phin &=& [17.00,  20.00] \\
\log \nx   &=& [10.25, 14.00]  \\
\log U_{\rm min} &=& [-6.5,  -4.5]
\end{array}
\end{equation} 
To limit the number of assumptions in the calculation, 
these integration limits span a larger range than suggested by the
observational limits discussed in Section~\ref{sect:param}.
The \phin\ and \nx\ integration limits were stepped in 0.25~decade intervals and 
the \Un\ limit in 0.5~decade intervals. 
The \phix\ and \nn\ limits were also varied, however the effect on the
total emission was small, because the \HI\ and \HeI\ lines have low emissivity in
this region of parameter space. The values of $\phix$ and $\nn$ were fixed
at $10^{23}$\cmsqs\ and $10^8$\cmcu, respectively. The change in predicted ratio
caused by varying these limits was less than 1\%. 

To visualise how the ratios change as the \phin, \nx\ and $\Un$
integration limits are varied, several flux ratios are plotted against each of
the integration limits. 
Figure~\ref{figure:varylims} plots several predicted ratios: \pa/\pb, \pc/\pb,
\HeI\ 1.083$\mu$/\pb, \Ha/\Hb\ and \HeI\ 1.083$\mu$/\HeI\ \ll5876 as a function
of  $\phin$, $\nx$ and $\Un$ integration limits. The three panels show ratios
plotted as a function of each of the integration limits over the range given in
Equation~\ref{eqn:intlimits}, while the  other integration limits are held
constant at fiducial values (given in the legend). 


\section{Application to data}\label{sect:data}

\begin{figure*}
\centering
\includegraphics[width=0.95\linewidth]{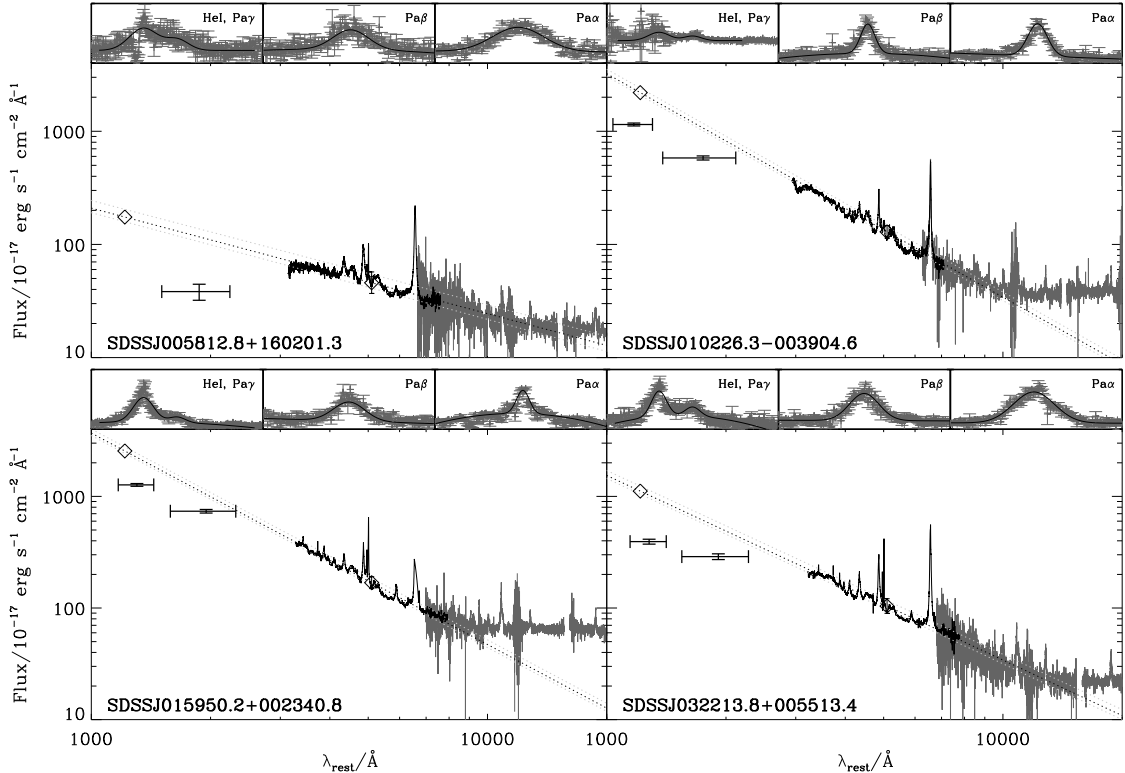}\\
\caption{\label{figure:data} 
Glikman NIR spectra (dark grey), scaled SDSS spectra (black), GALEX far and near UV
measurements (error bars). The extrapolated optical
continuum (dotted line) is also shown, with an estimate of the 1216\AA\ continuum flux marked
with a diamond. The insets show closeups of the \pc+\HeI\ 1.083$\mu$, \pb\ and \pa\ emission lines with best fit gaussians.
}
\end{figure*}

Very few quasars have been observed in the infrared with high signal-to-noise.
\citet{glikman2006} published a set of 27 low signal-to-noise quasar spectra,
that were presented as a composite quasar spectrum. Averaging quasar spectra is
complicated by several factors, which limits the accuracy of emission line
fluxes measured from the composite spectrum, see
Section~\ref{sect:discusscomposite} for details. 


Rather than using the composite spectrum, individual spectra with the
highest signal-to-noise ratios (SNR) from \citet{glikman2006} were analysed.
In practise our cutoff was $\mathrm{SNR} > 10$ per pixel in the continuum
between the \pa\ and \pb\ emission lines (excluding any regions of significant
atmospheric absorption).  Although there have been other NIR observations of
Seyferts \citep{landt2008}, only quasars were studied, as the subtraction of a
significant narrow component in Seyfert galaxies adds additional
uncertainty, which must be quantified in order to compare measured to predicted
line fluxes. 

\subsection{Measured broad emission line fluxes}

\tabletypesize{\small }
\begin{deluxetable*}{l rrrr rrrrr}
\tabletypesize{\small}
\tablecaption{
\label{table:measuredflux}
Measured NIR and optical line fluxes}
\tabletypesize{\tiny}
\tablehead{
         & \HeI\ 5876\AA &\Hc\  4341\AA & \Hb\ 4861\AA & \Ha\ 6563\AA & \HeI\ 1.083$\mu$m & \pc\ 1.094$\mu$m & \pb\ 1.282$\mu$m & \pa\ 1.876$\mu$m\\
Object  &$F$ ~~$ \varepsilon(F)$ &$F$ ~~$\varepsilon(F)$ & $F$ ~~$\varepsilon(F)$&$F$ ~~$\varepsilon(F)$&$F$ ~~$\varepsilon(F)$&$F$ ~~$\varepsilon(F)$&$F$ ~~$\varepsilon(F)$&$F$ ~~$\varepsilon(F)$}
\startdata
\objone     &347 10\%   & 1285 10\% & 4364 10\% &11446 10\% &1580 14\% &626 17\% & 1078 14\% & 1388 12\% \\
\objtwo     &630 11\%   & 2577 10\% & 6281 10\% &18003 10\% &2776 11\% & 1264 10\% & 1370 10\% & 1661 10\% \\
\objfive$^a$& 2452 10\% & 4259 10\% & 7506 10\% &13447 50\% &5222 10\% & 1296 11\% & 2100 11\% & 2772 11\% \\
\objseven   &682 10\%   & 3883 10\% & 7888 10\% &22781 10\% &2202 11\% & 1151 11\% & 1779 10\% & 2421 11\% \\
\enddata
\tablecomments{
Measured NIR and optical line fluxes ($10^{-17}$~erg$\,$s$^{-1}\,$cm$^{-2}$) with their associated measurement uncertainties. 
The optical lines: \Ha, \Hb, \Hc\ and \HeI\ \ll5876 were measured directly from the SDSS spectra, 
and have been scaled by $f_\mathrm{SDSS}$ from Table~\ref{table:matching} for comparison with the infrared lines.  
A 10\% systematic uncertainty was estimated on each emission line flux, predominantly due to use of Gaussian fitting (see text).
\objfive$^{a}$ is a ULIRG with an extremely asymmetric \Ha\ line; the 50\% error on \Ha\ 
accounts for the inadequacy of a Gaussian fit for this line. 
}
\end{deluxetable*}


Four quasar spectra from the \citet{glikman2006} sample have high enough
signal-to-noise ratios to measure the strongest Paschen emission line fluxes. 
For each emission line in the SDSS and \citeauthor{glikman2006} spectra, the
local continuum was measured over a wavelength range $\pm 200-500$\AA, centred
on the line's expected wavelength.  Linear fits to the continuum were performed
after removing the line itself and any 2$\sigma$ outliers. 
To measure line strengths, both gaussian fits and simple integrations of the
flux per wavelength bin
%
%
were performed for each line.  While gaussians are an imperfect match to the
line shapes, they minimise parameterisation while allowing for de-blending of
the \HeI\ and \pc\ lines. Integrating the flux per wavelength bin over $\pm 2$
FWHM reproduced the gaussian line fluxes within the errors. Summing over $\pm
3$ FWHM admits smaller outlying features and so tends to result in an
over-estimate of the line strength.  Measured line fluxes are shown in
Table~\ref{table:measuredflux} together with their 1-$\sigma$ measurement
uncertainties. The SDSS line strengths were scaled by $f_{\rm SDSS}$ to allow
direct comparison. 
%
%
An additional 10\% systematic error was conservatively added in quadrature to
measurement errors in all figures and calculations.

\subsection{Measured continuum fluxes}\label{sect:measure}

\tabletypesize{\scriptsize}
\begin{deluxetable*}{lc rr rr r rrr rr}
\tablecaption{
\label{table:matching}
Observables, flux-matching and estimates of the 1216\AA\ and 5100\AA\ continuum fluxes}
  \tabletypesize{\footnotesize}
\tablehead{
&  & \multicolumn{4}{c}{\galex}& & \multicolumn{4}{c}{SDSS}\\
\cline{3-6} \cline{8-11}
Object & $z$ & $S_\mathrm{FUV}$ & $S_\mathrm{NUV}$ & $E_{B-V}$  &$S_{1216}$ 
& & $S_{1216}$  & $S_{5100}$& $f_\mathrm{SDSS}$ &$\varepsilon(f_\mathrm{SDSS})$ 
}
\startdata
SDSSJ005812.8$+$160201.3  & 0.211 & ...   &  20.6 & 0.075 & ...     & & 110  &  42   & 0.729  & 1.2\% \\
SDSSJ010226.3$-$003904.6  & 0.295 & 880   & 431   & 0.036 & 2300  & & 2200 & 130   & 1.452  & 0.6\% \\
SDSSJ015950.2$+$002340.8  & 0.163 & 1020  & 580   & 0.029 & 2500  & & 2600 & 170   & 1.073  & 0.6\% \\
SDSSJ032213.8$+$005513.4  & 0.185 & 164   & 108   & 0.120 & 790   & & 1100 & 100   & 1.003  & 0.8\% \\
\enddata
\tablecomments{SDSS target name; redshift; GALEX raw observed fluxes ($10^{-17}$erg$\,$s$^{-1}\,$cm$^{-2}\,$\AA$^{-1}$) at
$1516\pm134$\AA\ (FUV) and $2267\pm366$\AA\ (NUV);
Galactic extinction color excess~\citep{schlegel98};
GALEX estimate of continuum flux at 1216\AA;
SDSS-extrapolated continuum flux at 1216\AA\ and measured continuum at 5100\AA;
flux factor to match SDSS to the NIR spectra and its uncertainty, $\epsilon$.
Note that \galex\ $S_\mathrm{FUV}$ and $S_\mathrm{NUV}$ fluxes have not been corrected for
extinction in this table.
}
\end{deluxetable*}

For each of the four objects in the sample, SDSS
spectra and GALEX ultraviolet flux measurements were obtained from archival
data. Colour excess values, $E(B-V)$ \citep{schlegel98} are given in
Table~\ref{table:matching}, and these were used to correct all the data for
Galactic extinction following~\citet{fitzpatrick99}.
Note that although the continuum fluxes inferred in this section have large uncertainties,
the inferred physical conditions do not depend on these values.
The measured continuum flux values are only used
in Section~\ref{section:ewsanitycheck} to check that the model produces
enough energy to be consistent with previous observations.

For each object, the continuum level of the SDSS spectrum was scaled to match
the NIR continuum across the region of overlap. The scaling factor for each
SDSS spectrum, denoted $f_{\rm SDSS}$, was typically measured to better than
2\% accuracy. The continuum flux at 5100\AA\ ($S_{5100}$) was measured from the
SDSS spectrum, which can be scaled to match the NIR data using $f_{\rm SDSS}$.
An extrapolation of the best linear fit to the optical continuum was made to
give an estimate of the rest-frame 1216\AA\ continuum flux ($S_{1216}$(SDSS)).
$S_{1216}$(SDSS) is likely to overestimate the true continuum flux at 1216\AA\
due to the turnover in the spectrum. An alternative estimate of the 1216\AA\
continuum flux was obtained from the GALEX far-ultraviolet (FUV) measurement.
Due to the shape, breadth, and variable position of \lya\ in the GALEX bandpass
for our sample, we cannot use the equivalent width of \lya\ alone to estimate
the 1216\AA\ continuum flux. Instead we convolve an appropriately redshifted
 composite quasar spectrum \citep{vdb+01} with the GALEX FUV bandpass to
estimate the continuum level.
%
%
The SDSS and GALEX estimates of the 1216\AA\ continuum flux level are in good
agreement, considering the significant uncertainty in extrapolating the
continuum. 
Figure~\ref{figure:data} shows the NIR spectra, scaled SDSS spectra and GALEX
photometry.
%
%
%


\section{Results}\label{section:results}

\begin{figure}
\centering
\includegraphics[width=0.97\linewidth]{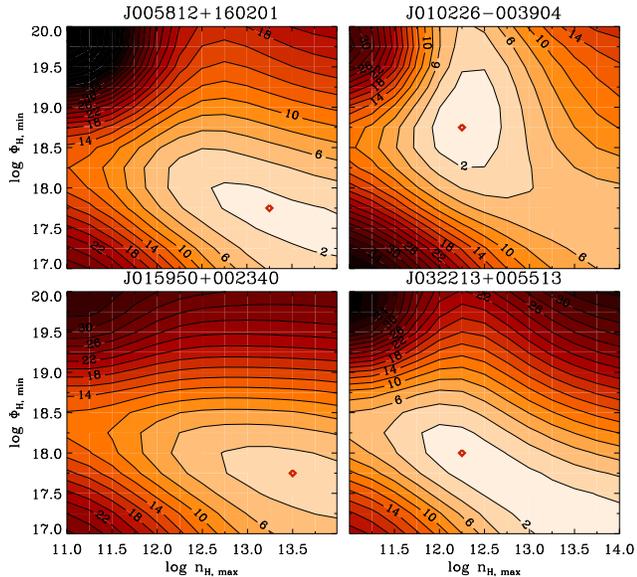}
\caption{\label{figure:chi2contour} 
Contours of reduced $\chi^2$ in the \phin--\nx\ plane for each of the four objects. 
Measured \pa/\pb, \pc/\pb\ and \Ha/\pb\ ratios were used to calculate $\chi^2$. 
In each panel, the location of the minimum $\chi^2$ is shown with a red diamond 
(see Table~\ref{table:ratiopb} for minimum values of reduced $\chi^2$). 
}
\end{figure}

The primary aim of this paper is to find a set of integration limits for the
hydrogen number density and hydrogen ionising flux (see
Equation~\ref{equation:loc}) that best fit the measured infrared hydrogen
emission line ratios for individual quasars.  
Since there are only a few measured infrared ratios, only a small number of
model free parameters can be fitted. Thus the aim of this analysis is to
illustrate the potential of using optical and infrared hydrogen and helium
emission lines as probes of the physical conditions within the BELR, rather
than constraining a definitive model. 

\subsection{Best-fit integration limits}\label{sect:resultsir}

A set of `best-fit' integration limits was found for each object using the
measured \pa/\pb, \pc/\pb\ and \Ha/\pb\ ratios. Ratios involving optical
emission lines were not used, as they suffer differential extinction from dust
and the amount of dust is unknown. 
There are only three independent infrared hydrogen ratios, so the model must
be limited to fitting to two parameters. Since the ratios are fairly constant
with the lower limit on the ionisation parameter, $\log \Un =-5.5$ is fixed,
and \phin\ and \nx\ are free parameters. 

The measured emission line ratios of four objects:  \objone, \objtwo, 
\objfive\ and \objseven\ were used to find 
the most likely set of \phin\ and \nx\ limits by minimising $\chi^2$:
\begin{equation}
\chi^2 = \sum_{i=1}^N \frac{(R_{{\rm obs},i} -R_{{\rm model},i} )^2}{ \sigma_{{\rm stat},i}^2 +  \sigma_{{\rm sys},i}^2},
\end{equation}
where $R_{{\rm obs}}$ and $R_{{\rm model}}$ are the observed and predicted
ratios, and $\sigma_{\rm stat}$ and $\sigma_{\rm
sys}$ are the statistical and systematic uncertainties on each measured ratio.
The sum is over each ratio used to find the best model: \pa/\pb, \pc/\pb\ and
\Ha/\pb. 
$R_{{\rm model}}$ is the predicted ratio for each combination of \phin\ and
\nx\ values calculated using Equation~\ref{equation:loc}, over the ranges given in Equation~\ref{eqn:intlimits}.
Table~\ref{table:ratiopb} shows the minimum $\chi^2$ value for each object.
Figure~\ref{figure:chi2contour} shows the $\chi^2$ contours for each object in
the \phin--\nx\ plane. This plot shows that the $\chi^2$ value is minimised where the 
density upper integration limit is high and the lower limit on the ionising flux is low. 

\begin{figure*}
\centering
\includegraphics[width=0.80\linewidth]{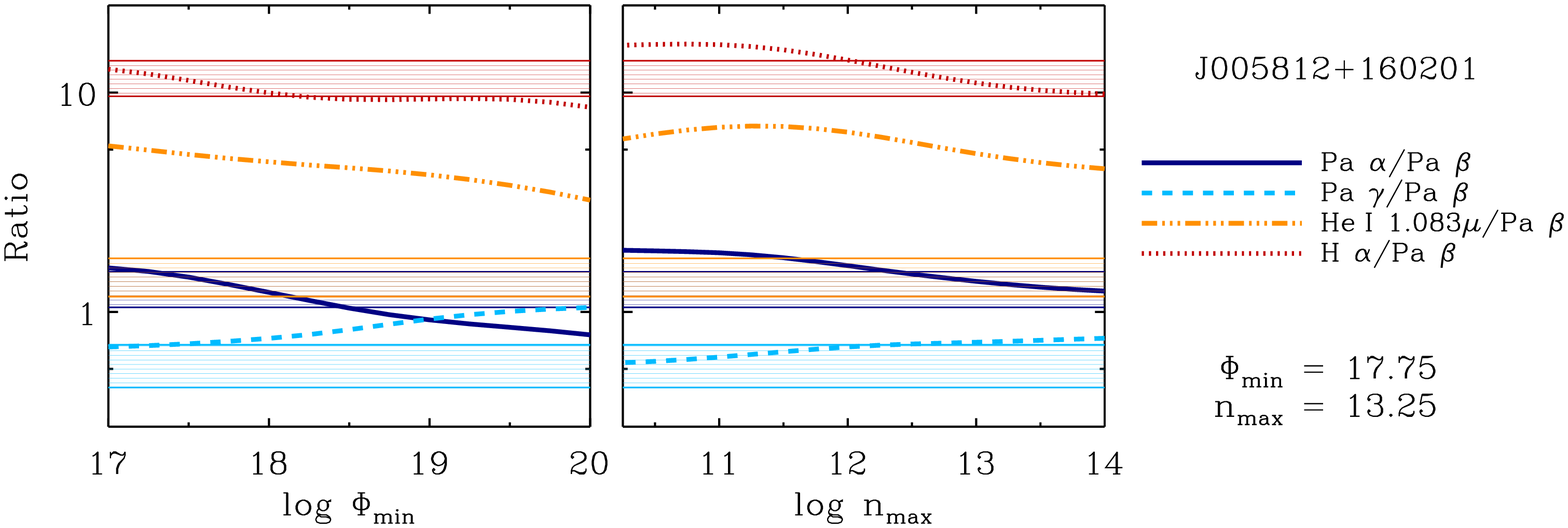}\\
\includegraphics[width=0.80\linewidth]{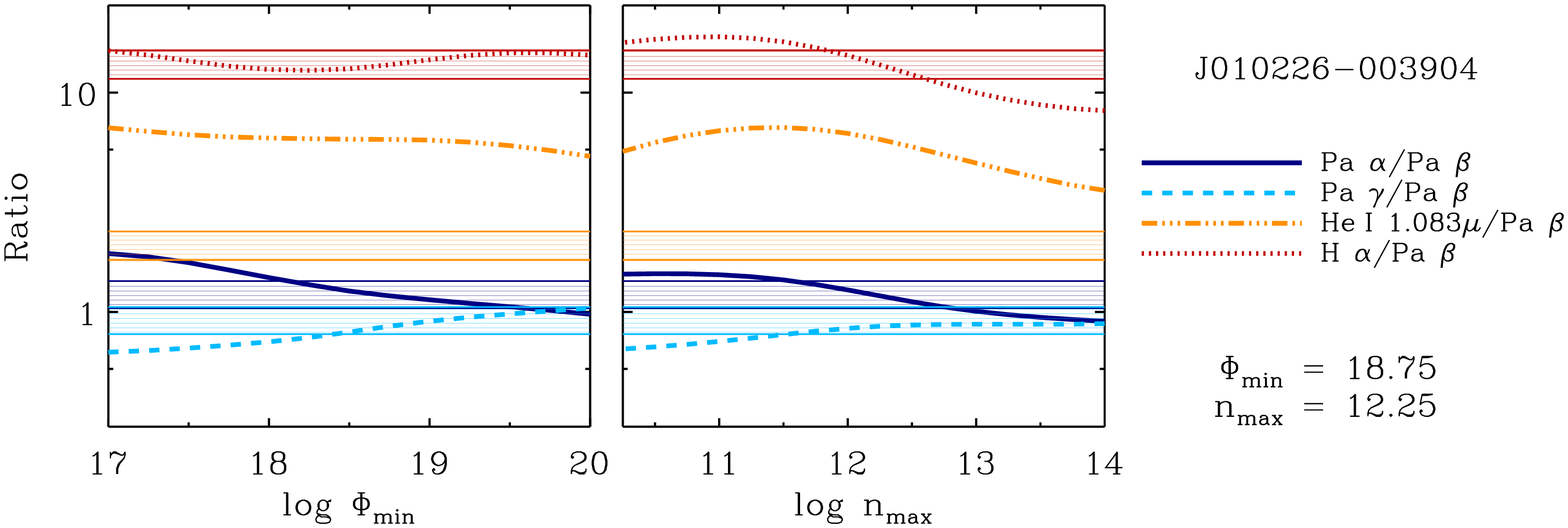}\\
\includegraphics[width=0.80\linewidth]{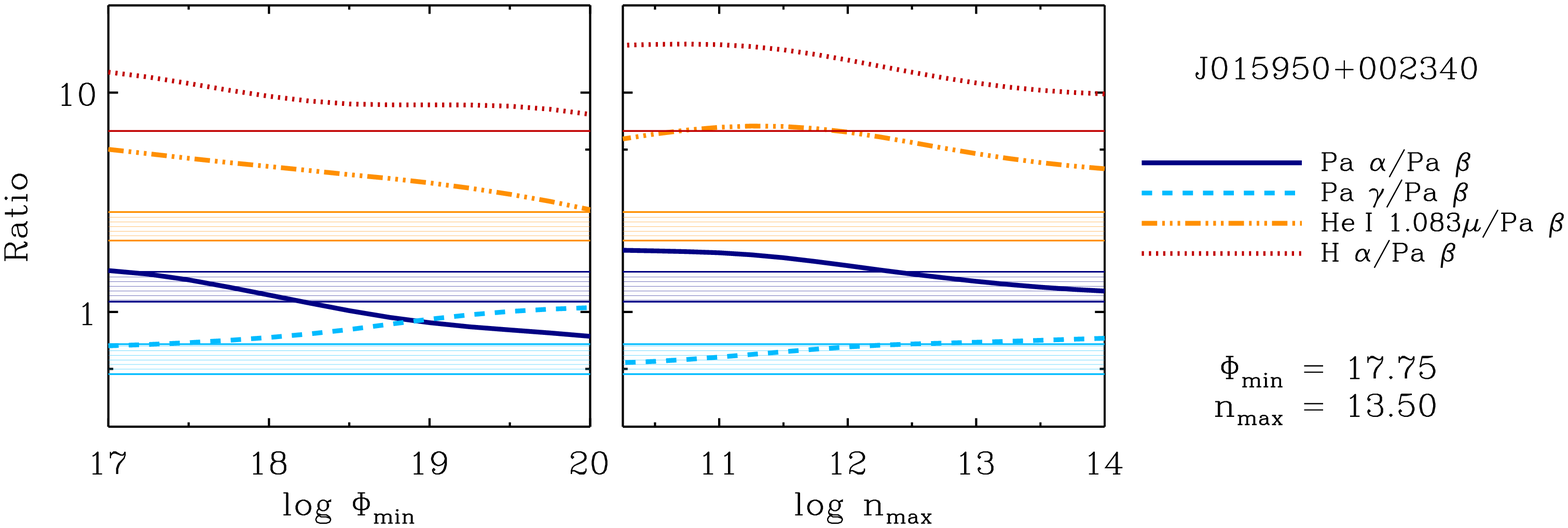}\\
\includegraphics[width=0.80\linewidth]{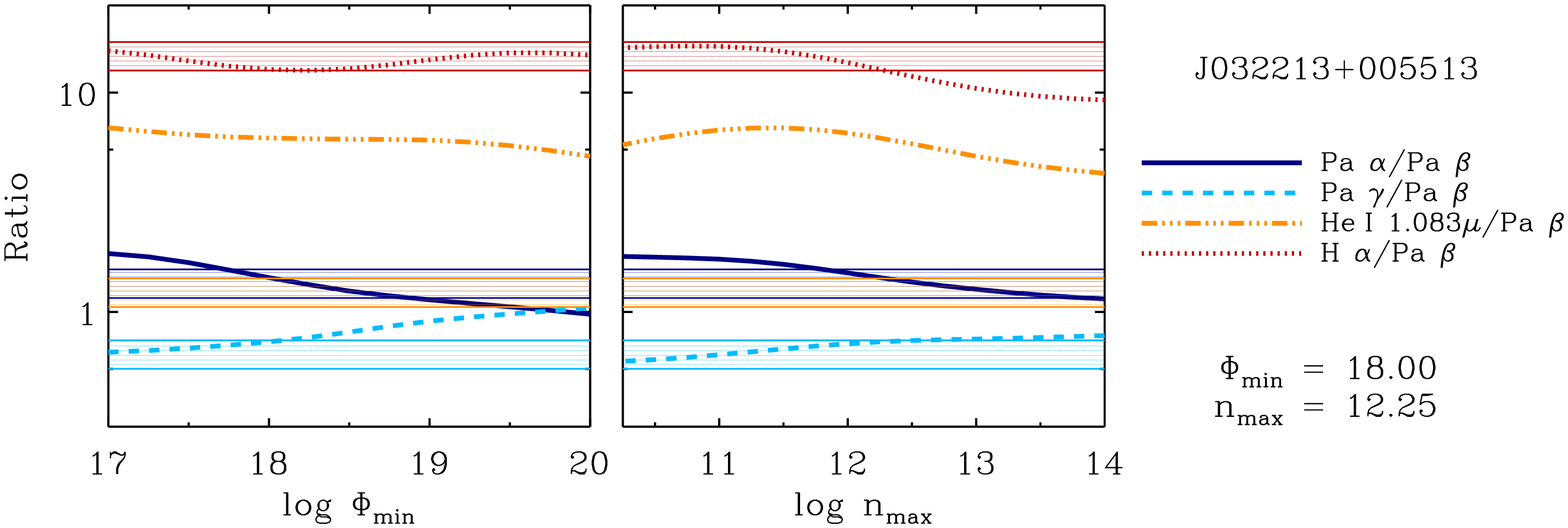}\\
\caption{\label{figure:glikbf} 
Measured ratios are plotted compared to the best-fit models. 
The solid, dashed, dot dashed and dotted lines show the simulated \pa/\pb,
\pc/\pb, \HeI\ 1.083$\mu$/\pb\ and \Ha/\pb\ ratios, respectively. 
The measured values are shown as a band with 1$\sigma$
uncertainties in a corresponding colour. 
For legibility, the very large \Ha\ measurement uncertainty is not 
shown for \objfive, although it was used in the best-fit calculation. 
%
The parameters that produce the best fit for each object have been plotted and the 
values are given in the legend for each object. 
Note that a log scale has been used on the y-axis, so that the range of ratios can 
easily be seen.  
}
\end{figure*}

Figure~\ref{figure:glikbf} shows the best-fit simulated ratios as a function of
the two free parameters: \nx\ and \phin. The best-fit \nx\ and \phin\ limits
are given in the legend for each object. The \nn\ and \phix\ integration
limits were fixed at fiducial values given in Section~\ref{sect:limits}. The
measured infrared ratios are over-plotted as a band, indicating a 1$\sigma$
uncertainty. 
Note that the \HeI\ 1.083$\mu$/\pb\ ratio was not used as a constraint, as this
ratio predicts \phin\ and \nx\ values that are systematically offset from those
given by the hydrogen line ratios, as shown in Figure~\ref{figure:glikbf}.


\subsection{Predicted Balmer ratios}

The limits on \nx\ and \phin\ that were calculated by analysing the NIR emission lines 
were then used to predict \Ha/\Hb\ and \Hc/\Hb\ ratios. 
The predicted ratios are shown with the measured ratios in
Table~\ref{table:ratiopb}. Two of the four objects are in the \citet{dong2008}
sample, that selected AGNs with blue optical continua in order to minimise the
effect of dust extinction.
If blue continua do indicate less dust along the
line of sight, the simulated ratios should be in good agreement with the
measured ratios for these objects. This is true for the two objects listed, 
however, a much larger sample with a larger number of ratios studied is required to
test this hypothesis. \objfive\ also has a blue continuum, however simulated
values cannot be compared to measurements, due to the poor fit to the \Ha\
line.

For all objects, the predicted Balmer decrements are slightly larger than the measured
values, which is the opposite effect to dust. However, the predicted values
are within measurement uncertainties, which may indicate that there is little internal 
reddening in these objects. 
The predicted \Hc/\Hb\ ratios are consistent with the measured data. The results for 
each object are summarised briefly in Section~\ref{section:objresults}.

\tabletypesize{\small}
\begin{deluxetable*}{lc cc cc cc cc cc cc}
\tablecaption{\label{table:ratiopb}
Measured and predicted ratios}
  \tabletypesize{\small}
\tablehead{ 
                 &           & \multicolumn{6}{c}{Used for fit} & \multicolumn{6}{c}{Not used for fit}\\
\cline{4-7} \cline{10-13}
\colhead{Object} &  $\chi^2$ &  \multicolumn{2}{c}{\pa/\pb} & \multicolumn{2}{c}{\pc/\pb} & \multicolumn{2}{c}{\Ha/\pb}& 
  \multicolumn{2}{c}{\Ha/\Hb}& \multicolumn{2}{c}{\Hc/\Hb}  & \multicolumn{2}{c}{\HeI\ 10830/\HeI\ 5876}  \\
   &       &      m        & p     &        m      &    p  &     m        &     p  &       m      &     p  &      m        &   p  &          m     &   p   
}   
\startdata
\objone      &1.5 &   1.29$\pm$0.2&   1.34&   0.58$\pm$0.1&   0.74&  10.6$\pm$2.2 &  10.6&   2.62$\pm$0.4&   3.04&   0.29$\pm$0.1&   0.53&   4.54$\pm$0.8&   3.09 \\ 
\objtwo$^d$  &0.27&   1.21$\pm$0.2&   1.18&   0.92$\pm$0.1&   0.86&   13.1$\pm$2.0&  13.4&   2.87$\pm$0.4&   3.18&   0.41$\pm$0.1&   0.46&   4.41$\pm$0.7&   2.42 \\ 
\objfive     &  3.1&   1.32$\pm$0.2&   1.30&   0.62$\pm$0.1&   0.74&   6.40$\pm$3.0&  10.3&   1.79$\pm$0.9&   2.90&   0.57$\pm$0.1&   0.55&   2.13$\pm$0.3&   2.96 \\ 
\objseven$^d$& 0.90&   1.36$\pm$0.2&   1.44&   0.65$\pm$0.1&   0.73&   12.8$\pm$2.2&  12.8&   2.89$\pm$0.4&   3.60&   0.49$\pm$0.1&   0.45&   3.23$\pm$0.5&   3.23 \\ 
\enddata
\tablecomments{Measured (m) and predicted (p) broad emission line flux ratios. 
Using the best-fit parameters discussed in 
Section~\ref{sect:resultsir}, the \Ha/\Hb, \Hc/\Hb\ and \HeI\ 1.083\micron/\HeI\ \ll5876 ratios 
were predicted for each object. The predicted ratios are
compared to the measured SDSS ratios. 
A 10\% systematic was included on each measured flux.
Objects that are in the \citet{dong2008} sample are denoted by $d$. 
 }
\end{deluxetable*}

\subsection{The predicted \lya/\Hb\ and UV line ratios}

Although UV emission lines have not been measured for these objects, a
selection of predicted UV emission line ratios are presented in this section. 
Typical measured strengths of stronger UV lines were compared to the predicted
values to check the consistency of our spectral model derived from the
measured hydrogen spectrum.  
%
%
We note that the UV metal emission line strengths depend more on the
shape of the incident ionising continuum (and to varying degrees the gas
metallicity) than do the Balmer and Paschen hydrogen lines under study
\citep[\eg, see][for a discussion]{korista1998}.  The predicted UV emission
line strengths reported here are those that result from the best-fit in the
physical parameters (\nx, \phin) determined from the three hydrogen
emission line ratios (as noted in Table~\ref{table:ratiopb}) in each of the four AGN.



%
Predicted \lya/\Hb\ ratios for each object are listed in Table~\ref{table:predictuv}.
Most of the \lya/\Hb\ predicted ratios are lower than the simple, classical
photoionisation prediction \citep{book:oster}. This is because of the inclusion
of high density gas. \lya\ thermalises at high densities, whereas the Balmer
lines continue to emit efficiently, causing the predicted \lya/\Hb\ ratio to
decrease with the inclusion of high \nh\ gas. Nevertheless, these ratios remain
larger than many of those measured in AGN, and this remains an unsolved
problem. However, uncorrected extinction due to grains extrinsic to the Milky
Way may explain much of the remaining discrepancies \citep[see][]{netzer1995}.
%

A selection of predicted UV metal lines (relative to \lya) are listed in
Table~\ref{table:predictuv}. These lines were chosen as they are relatively
strong, and they span a large range in ionisation parameter. 
%
%
The \OVI, \CIV\ and \MgII\ emission lines are also the most important
coolants within their ionisation zones (excepting emission by \FeII). 
As these lines are major coolants, the line strengths are not
very sensitive to the gas metallicity, due to the thermostatic effects of such
lines. 
The predicted \CIV\ values are somewhat higher than the ratio measured from the
\citet{vdb+01} composite spectrum, however, the predicted \CIV/\lya\ ratio is
strongly dependent on the hardness of the ionising continuum. 
%
%
The observed value shows significant variation; for
example, the \CIV/\lya\ ratio is 0.62 in the \citet{zheng1997} composite, and is 
closer to 1 in Seyfert~1 spectra (\citealp{KG2000}; see also \citealp{osmer1994}). 
Although additional free parameters, such as metallicity and the shape of the
ionising continuum, have not been considered here, the predicted strengths of
the UV metal lines relative to \lya\ are generally consistent with typical
measured values.



\begin{deluxetable*}{l c cc c cc c}
\centering
\tablecaption{\label{table:predictuv}
Predicted \lya/\Hb\ and ultraviolet line ratios}
\tabletypesize{\small} 
\tablehead{Object  & \lya/\Hb &\OVI$+$\lyb/\lya &   \CIV/\lya & \MgII/\lya        
}
\startdata
\objone      & 19 & 0.14 &  0.33& 	0.22 \\ 
\objtwo      & 40 & 0.24 &  0.40&  0.09 \\ 
\objfive     & 17 & 0.14 &  0.33&  0.22 \\ 
\objseven    & 31 & 0.16 &  0.35&  0.16 \\ 
\hline
measured$^a$ & 12 & 0.10 &  0.25&  0.14 \\
\enddata
\tablecomments{
Predicted \lya/\Hb, and \OVI\ \ll1035$+$\lyb, \CIV\ \ll1549 and 
\MgII\ \ll2800 line ratios relative to \lya. 
The line ratios were predicted using the best-fit parameters derived 
in Section~\ref{sect:resultsir}.
\newline\noindent $^a$ The measured line ratios are from the \citet{vdb+01} composite spectrum.
}
\end{deluxetable*}


\subsection{Do the models produce enough energy?}\label{section:ewsanitycheck}

A simple check, albeit with a substantial uncertainty, can establish whether the best-fit 
models produce enough energy to reproduce the observed total integrated
line flux. The largest uncertainty in comparing measured equivalent width
values to predicted line strengths is knowing which continuum flux to compare
the line flux to.
%
%
The NIR continuum is a poor indicator of the incident ionising continuum (\ie\
photons with wavelengths shorter than 912\AA), and is contaminated with hot
thermal dust emission as well as that of the host galaxy.
To test whether the models produce enough energy, the predicted and measured \pb\
equivalent widths were compared, relative to two different points in the
continuum: 1216\AA\ and 5100\AA. 
Note that the predicted equivalent width depends on the integrated covering
factor.  In this section, the integrated covering factor was calculated, such
that the simulated equivalent width of \lya\ is 100\AA. This is a typical
measured value for \ew(\lya), and was chosen to ensure that the model produces
enough energy to be consistent with the observed value. For each object, this
value of \fc\ was used to calculate all equivalent width values presented in
this section. 

The equivalent width relative to the continuum flux at 1216\AA\ is often chosen 
because it is easily observable and close to the ionising continuum. Contaminants 
such as light from the host galaxy and dusty torus are negligible in this region.   
The measured value was estimated using both the \galex\ and SDSS UV continuum flux
values ($S_{1216}$) listed in Table~\ref{table:matching}. 
The observed rest-frame equivalent width of \pb\ relative to the incident
continuum at 1216\AA\ (in \AA) is given by
\begin{equation} W_{1216}({\rm \pb}) = \frac{F({\rm \pb})}{S_{1216}}\times
\frac{1}{(1+z)}, 
\label{equation:ewpb}
\end{equation} 
where $S_{1216}$ is the measured incident continuum flux at 1216\AA\ and
$F({\rm \pb})$ is the total integrated emission line flux, given in
Table~\ref{table:measuredflux}.  
The dominant measurement uncertainties are quasar variability and scaling the
UV measurement to the NIR. There is an additional uncertainty from dust attenuation. 
The measured $S_{1216}$ values show a large variance between the \galex\ and \sdss\ 
estimates, therefore this value is estimated to be accurate within a factor of 2. 

The measured \pb\ equivalent width value relative to the
incident continuum at 5100\AA\ (\ewopt(\pb)) was also calculated. At this
wavelength the continuum flux is directly measurable from the scaled SDSS
spectrum (unlike the SDSS continuum at 1216\AA, which is an extrapolated
estimate). 
However, the 5100\AA\ continuum is much further from the ionising continuum and
host galaxy contamination is not negligible at this wavelength, making it a
poorer proxy for the ionising continuum.  
There may also be some thermal emission from
the BELR clouds and reflected light from the dusty torus contributing to the
measured continuum flux at this wavelength.  The calculation of \ewopt(\pb) is
analogous to the calculation of \ew(\pb) given in Equation~\ref{equation:ewpb}. 
Given the large uncertainties in scaling the continuum measurements to the NIR,
quasar variability and host galaxy contamination in the $S_{5100}$ measurement,
the measured values listed in Table~\ref{table:predictew} should only be
considered accurate to within a factor of~2.

Using the best-fit \nx\ and \phin\ integration limits determined in
Section~\ref{sect:resultsir}, the simulated
\ew(\pb) and \ewopt(\pb) values were calculated for each object. 
Both measures of equivalent width can be output by \cloudy\ directly.  (Note
that these estimates assume the standard \cloudy\ continuum described in
Section~\ref{section:grid}.) 
%
To compare the simulated equivalent width values to the measured values, 
predicted \ew(\pb) and \ewopt(\pb) values
must be scaled by the integrated cloud covering fraction (\fc).
For each object, the integrated covering fraction required to produce \ew(\lya)=100\AA\ (a
typical value) was computed for the best-fit parameters.  
This value of \fc\ was then used to scale the predicted equivalent width values.

The measured and predicted \ew(\pb) and \ewopt(\pb) values, and the value of
\fc\ are listed in Table~\ref{table:predictew}.
As shown in this table, the computed integrated covering fraction required to
produce \ew(\lya)=100\AA\ for each object is not unrealistically large.  The
measured and predicted \ew(\pb) values are also in good agreement (with the
exception of \objone, which has a shallower UV continuum slope than the \cloudy\
AGN continuum).  Therefore Table~\ref{table:predictew} shows that the best-fit
models produce enough energy to be consistent with both the measured \pb\ and
previously observed \lya\ equivalent width values. 



\begin{deluxetable}{l ccc c cc c }
\tablecaption{\label{table:predictew}
Predicted rest frame equivalent width of \pb }
\tablehead{ 
\colhead{Object} &    \multicolumn{3}{c}{\ew(\pb)} & & \multicolumn{2}{c}{\ewopt(\pb)} & \fc \\
\cline{2-4} \cline{6-7}
   &      {\sc galex}        &         {\sc sdss}      &    p  &    & {\sc sdss}        &     p  & p
}   
\startdata
\objone      & ...   & 10 &  1.5&   &  30 &   12   &  0.31\\ 
\objtwo$^d$  &  0.3  &0.3& 0.57&   &   6 &   4.5  &  0.41\\ 
\objfive     &  0.7  &0.6&  1.7&   &   10&   13.3 &  0.32\\ 
\objseven$^d$&  2    &1  & 0.89&   &   10&   7.1  &  0.30\\ 
\enddata
\tablecomments{Measured and predicted (p) \pb\ equivalent width values (in \AA) normalised to two different points 
in the continuum: 1216\AA\ (SDSS and \galex) and 5100\AA\ (SDSS only). 
The predicted equivalent width values are calculated using the best-fit parameters and
were scaled using an integrated covering fraction that gives a predicted \lya\ equivalent width of 100\AA.
The value of \fc\ required to produce \ew(\lya)=100\AA\ is also given.
Uncertainties on the measured equivalent width values are large, and these values are estimated to be accurate to a factor of $\sim$2.
}
\end{deluxetable}

\subsection{Results for each object}\label{section:objresults}

A brief description of the results for each object is given below.

{\bf \objone}: the model is a good fit to the data, although the predicted
\Hc/\Hb\ ratio is 2$\sigma$ larger than the measured ratio. The measured
\Hc/\Hb\ ratio is low compared to the other objects. The predicted total
equivalent width barely produces enough energy to be consistent with the
observations. The low predicted equivalent width of \pb\ is at least partially
due the assumed \cloudy\ continuum, which is a poor approximation in this case,
as the measured UV spectrum is much shallower than the \cloudy\ continuum.  The
measured \ew(\pb) is high compared to the other objects, again because the
continuum slope is very shallow. 


{\bf \objtwo}: the model is a good fit to the data. This object has a strange BAL-like
feature in \HeI\ 1.083$\mu$, probably due to the atmosphere. However, this feature
could be real, similar to that reported by \citet{leighly2011}. The predicted
\phin\ value is quite high, which results from the large measured \pc/\pb\
ratio.

{\bf \objfive}: the model is a fair fit to the data, as the
reduced $\chi^2$ value is 3.1. This source, also known as Mrk~1014, is an
ultra-luminous infrared galaxy (ULIRG) \citep[see][and references
therein]{armus2004}. The \Ha\ emission line is highly asymmetric and is not
well fit by a Gaussian, making comparisons to simulations difficult. The high
density predicted could be a  consequence of the underestimation of the \Ha\
flux.  
Note that the measured \Ha/\pb\ and \Ha/\Hb\ ratios are low compared to the
other objects. Also note that the optical and NIR \HeI\ line flux (relative to
the hydrogen line flux) is high compared to the other objects.  


{\bf \objseven}: the model produced a good fit to the emission line
ratios, although the predicted \Ha/\Hb\ ratio is $2\sigma$ above the measured value.


\section{Discussion}\label{section:discussion}

The predictions of \cloudy\ LOC models have been tested by comparing simulated
ratios to measured near-infrared hydrogen emission line ratios. Near-infrared
hydrogen emission lines were chosen to better understand hydrogen emission from
the BELR, and because near-infrared ratios are insensitive to uncertainties in
the line-of-sight extinction due to the presence of dust. 
The signal-to-noise of the published spectra is limited, so only four quasars
were considered suitable. Two questions were considered: \par (i) Is a good fit
to the observed line ratios possible using the LOC model?  and \par (ii) What
are the integration limits for the best-fit LOC models?
\newline\noindent Due to the limited number of independent line ratios, only two integration
limits in the LOC model could be tested.


\subsection{Can a simple LOC model fit these observations?}

In three of the four cases, a good fit to the hydrogen emission lines is
obtained by the LOC model. In the fourth case the fit is fair. In addition, the
shape of the $\chi^2$ contour distributions in the density-flux plane is
similar for all objects.  
High \phin\ limits are generally highly disfavoured. This suggests that $\log
\phin<18$, and that low incident ionising fluxes are required to reproduce
observed hydrogen emission line flux ratios. The determined limits on \phin\
are also consistent with the dust sublimation radius \citep{nenkova2008b}.
This is in agreement with several recent studies that suggest the Balmer
lines and \MgII\  in large part come from the outer BELR and near or at the dust
sublimation radius \citep{zhu2009,czerny2011,mor2012}.

Figure~\ref{figure:chi2contour} shows that lower \nx\ values in
combination with low \phin\ values are also highly disfavoured (\ie\ $\chi^2$
values are high in the lower left corner of the plot), meaning that BELRs with
a low high-density cutoff and low low-incident ionising flux cutoff are
unlikely solutions.
The minimum of the $\chi^2$ contours tend to lie on a diagonal in the
\nx--\phin\ plane: if \nx\ is larger, then \phin\ is smaller. If the BELR
includes regions where the incident ionising flux is very low, then gas with a
very high number density is required to reproduce the observed emission line
ratios. This solution is quite unlikely though, as dust grains condense at low
\phih, destroying the line emission.  The validity of the inferred limits
is uncertain. A larger sample, with higher signal-to-noise is required
to test these results.

The predicted \Ha/\Hb\ and \Hc/\Hb\ ratios are generally in very good agreement
with the measured ratios, as shown in Table~\ref{table:ratiopb}.
However, the measured \HeI\ 1.083$\mu$/\pb\ ratio does not match the simple LOC
model described here.  
We also find that the predicted \HeI\ 5876/\Hb\ line ratio is significantly
higher than the measured values. 
The simulated values are typically a factor of $\sim$2 greater than the
measured values. This offset could be caused by the simplicity of the covering
fractions assumed in this model. The shape of the \ew(\HeI\ 1.083$\mu$)
contours is quite different to the Balmer and Paschen lines, making ratios
involving \HeI\ 1.083$\mu$ more sensitive to the power laws assumed in
Equation~\ref{equation:loc1}.  Although the slope of the power laws are not
expected to deviate significantly from $-1$, refinement of the models using
different power law distributions in $f(r)$ and $g(\nh)$ will be the subject of
future studies. 
The predicted overestimate of both the \HeI\ \ll5876/\Hb\  and 1.083$\mu$/\pb\
line ratios could also indicate that \cloudy\ overpredicts the \HeI\ emission spectrum relative to H.






\subsection{Dust}

The amount of attenuation from internal dust (\ie\ dust within the host galaxy)
is unknown. The Balmer decrement is the most widely measured ratio is AGN
spectroscopy \citep[\eg][]{osterbrock1977,groves2012}, and has long been used
as an indicator of the amount of extinction \citep[\eg][]{berman1936}.
\citet{dong2008} measured the Balmer decrement for a sample of 446 Seyfert~1s
and QSOs with blue continua, which were selected to minimise the effect of dust
extinction. They found only a small intrinsic scatter in the measured
ratio. 
The small intrinsic scatter in the \citet{dong2008} measurement suggests that
the intrinsic Balmer decrement could be relatively consistent between objects,
and the observed variance is due to different amounts of dust along the
line-of-sight within the host galaxy. However, this result relies on the slope
of the continuum being a dust indicator \citep[see][for a
discussion]{davis2007}.

%

The continua of our subset of quasars are reasonably blue, with the exception
of \objone. If dust attenuation caused reddening of the continuum in this case,
a larger observed \Ha/\Hb\ would be expected. Unfortunately it is not possible
to test this assertion with such a limited sample size.
The presence of uncorrected extinction due to dust along the line of sight
would also increase the observed He I 1.083μ/He I λ5876 ratio above that of the
predicted one (see Table~\ref{table:ratiopb}). 


With better quality spectra, an estimate of the amount of dust attenuation
could be made. The \lya/\Hb\ problem could be at least partially explained by
internal extinction \citep{shuder1979} and constraining the amount of dust
attenuation is important to resolve this matter.

\subsection{Composite quasar spectra}\label{sect:discusscomposite}

The value of the \pa/\pb\ line ratio measured from the \citet{glikman2006} composite
spectrum was 0.64$\pm$0.01. However, this value is much lower than the values
measured from individual objects, which are consistently above 1.0 (see
Table~\ref{table:ratiopb}). 
Other observed \pa/\pb\ emission line ratios in quasars are consistent with this result
\citep[\eg][who measured a \pa/\pb\ ratio of 1.05$\pm$0.2 in a $z=3.91$
quasar]{soifer2004}. Ratios of \pa/\pb\ that are above 1.0 have also been
measured in lower luminosity AGN \citep{landt2008}. 

While it is possible that systematic effects cause objects with larger \pa/\pb\
ratios to be preferentially selected in our sample, a bias caused by the
composite methodology is a more likely explanation. 
Composite spectra are useful for describing generic properties of a sample and
for identifying emission lines that are too weak to be observed in individual
spectra. However, care must be taken in ascribing physical significance to
individual features of composite spectra. 
%
%
It is not clear whether composite spectra are useful for investigating detailed
physical properties of the line emitting gas. 
Emission line ratios in the composite spectrum can also be affected by: poor
signal-to-noise ratios of  individual spectra, and wavelength ranges where
there are few objects contributing to the final spectrum.  See
\citet{sulentic2002} for a discussion of how emission line ratios are affected,
and an alternative method for producing composite spectra.

\subsection{Caveats}

The NIR spectroscopy was not simultaneous with the optical and UV measurements.
As quasars are intrinsically variable, this will introduce a systematic
uncertainty. Even after continuum matching the line fluxes will alter, this
effect is known as the intrinsic Baldwin effect \citep[\eg][]{pogge1992}.
There is often a considerable difference between \galex\ and \sdss\ $S_{1216}$
values. The objects in Table~\ref{table:matching} \sdss\  are
all observed during 2000. NIR observations are from September 2003 to September 2004 
and \galex\ observations are in
2003 (and one in 2005). 

Fitting the emission lines with Gaussians introduces another systematic
uncertainty.  \citet{glikman2006} point out that Gaussians do not fit the broad
wings well, however Gaussians generally fit the bulk of the broad emission line
flux, while requiring only a small number of free parameters. 
%
%
The estimated contribution from the narrow component is less than
5\%. \citet{dong2008} deconvolved the emission lines into both a broad and narrow
component, and in each case the equivalent width of the narrow component is only a few
percent of the broad component. 
\FeII\ contamination was not considered. Fitting the \FeII\ contamination would
introduce additional errors, given the low signal-to-noise of the data.

Fixing the lower limit on the ionisation parameter, $\Un$ is not ideal,
however, the ratios considered are insensitive to $\un$ and with only a few
measured ratios, the model must be limited in complexity. The $\un$ limit would
be better determined with the \CaII\ infrared triplet (\Caxyz) or other
$\un$-sensitive ratios \citep[see][for a discussion]{ruff2011b}.  Single point
solutions in the density-flux plane were not considered for reasons discussed
in Section~\ref{sect:param}.

Since the Balmer and Paschen lines are optically thick, the geometry will
affect the predicted emission. If the BELR geometry is such that emission
from the `back' of the clouds is not observed, the predicted
emission line ratios will change \citep{ferland1992,obrien1994,ruff2011b}.


Any broad emission line flux measurement is from a snap-shot in time. The
ratios will vary with time as the continuum luminosity varies, and this will be
dependent on the distribution of the gas -- both in geometry and physical state. 
%
%
Comparing measured flux ratios can be problematic if the source is highly 
variable and lines are formed at very different radii. The recombination lines
are emitted relatively efficiently over a broad range of physical conditions,
and 
%
%
their continuum reprocessing efficiencies are largely dependent on the incident
flux, and therefore the distance from the ionising source, due to optical
depths in the excited states \citep{korista1997_atlas,ruff2011b}.
This manifests in different measured reverberation mapping signals and line 
profile variability \citep[\eg][]{sergeev2001,kol2010}. For
example, not only do the Balmer lines show different measured time lags, but
the line responsivity is observed to decrease from line centre to line wing
\citep[see][for a discussion]{KG2004}. 
This could be corrected for if the distribution of broad emission line gas
could be determined using high quality spectral and temporal data in conjunction
with photoionisation models \citep{HKG2003}.

\section{Summary}\label{section:summary}

We have demonstrated a new diagnostic for the density and ionisation state of
the hydrogen broad line gas in quasars.  The results should be seen as
preliminary, being based on fairly low SNR (SNR $\simeq10$ per pixel
in the continuum) data for a small number of objects.  The technique is
promising because the hydrogen line ratios depend strongly on the physical
conditions, but weakly on metallicity, column density and slope of the ionising
continuum. The ionising flux is a good measure of distance from the 
source, and the technique offers an alternative route to
measuring the structure of the quasar central engine, in addition to the
physical state of the broad line gas.

Despite the simplicity of the model, broad NIR hydrogen emission line ratios
can be reproduced by an LOC integration.  All hydrogen ratios predicted,
including the Balmer decrement, were consistent with those observed.  Low
incident ionising flux values ($\phih<10^{18}$\cmsqs) are required to
reproduce the observed hydrogen emission line ratios. This predicted lower
limit on \phih\ gives an outer radius of the BELR that is consistent with the radius of dust
sublimation calculated by \citet{nenkova2008b}. 

This model predicts that high number density gas ($\nh>10^{12}$\cmcu) is required to 
reproduce the observed hydrogen emission line ratios. The upper limit on
the gas number density is high compared to previous calculations, however, all
predictions stated here should be considered estimates, because of the
simplicity of the model and limitations in the quality and number of spectra.

With a larger sample of high signal-to-noise quasar spectra in the NIR, 
higher order lines in the Paschen series could also be observed. 
More constraints on the parameter space and other free parameters will be included
in the model.

\acknowledgments 

AJR acknowledges the support of an Australian Postgraduate Award.
We are grateful to Gary Ferland and collaborators for freely distributing and
maintaining the photoionisation code, \cloudy.  We thank Eilat
\citeauthor{glikman2006} for making their near-infrared spectra publicly
available. 
This research has made use of the NASA/IPAC Extragalactic Database (NED) which
is operated by the Jet Propulsion Laboratory, California Institute of
Technology, under contract with the National Aeronautics and Space
Administration.


The Infrared Telescope Facility (IRTF) is operated by the University of Hawaii
under Cooperative Agreement no. NNX-08AE38A with the National Aeronautics and
Space Administration, Science Mission Directorate, Planetary Astronomy Program.

Funding for the SDSS and SDSS-II has been provided by the Alfred P. Sloan
Foundation, the Participating Institutions, the National Science Foundation,
the U.S. Department of Energy, the National Aeronautics and Space
Administration, the Japanese Monbukagakusho, the Max Planck Society, and the
Higher Education Funding Council for England. The SDSS Web Site is
http://www.sdss.org/.  The SDSS is managed by the Astrophysical Research
Consortium for the Participating Institutions. The Participating Institutions
are the American Museum of Natural History, Astrophysical Institute Potsdam,
University of Basel, University of Cambridge, Case Western Reserve University,
University of Chicago, Drexel University, Fermilab, the Institute for Advanced
Study, the Japan Participation Group, Johns Hopkins University, the Joint
Institute for Nuclear Astrophysics, the Kavli Institute for Particle
Astrophysics and Cosmology, the Korean Scientist Group, the Chinese Academy of
Sciences (LAMOST), Los Alamos National Laboratory, the Max-Planck-Institute for
Astronomy (MPIA), the Max-Planck-Institute for Astrophysics (MPA), New Mexico
State University, Ohio State University, University of Pittsburgh, University
of Portsmouth, Princeton University, the United States Naval Observatory, and
the University of Washington.

GALEX is operated for NASA by the California Institute of Technology under NASA
contract NAS5-98034

\bibliographystyle{apj}


\end{document}